\documentclass[journal]{IEEEtran}
\ifCLASSINFOpdf
\else
\fi
\usepackage{color}
\usepackage{graphicx}
\usepackage{amsmath}
\usepackage{tabularx}
\usepackage{multirow}
\usepackage{booktabs}
\usepackage{cite}
\usepackage{datetime}

\newcommand{\sci}[3][10]{{#2}\times {#1}^{#3}}

\begin{document}
%
\title{Towards Simulation and Risk Assessment of Weather-Related Cascading Outages}
%
%
%

\author{Rui~Yao,~\IEEEmembership{Member,~IEEE,}~and~Kai~Sun,~\IEEEmembership{Senior~Member,~IEEE}
\thanks{This work was supported by the CURENT Engineering Research Center.}
\thanks{R. Yao and K. Sun are with the Department of EECS, the University of Tennessee, Knoxville, TN 37996, USA (emails: yaorui.thu@gmail.com, kaisun@utk.edu).}}

%
%

\markboth{
}%
{Shell \MakeLowercase{\textit{et al.}}: Towards Simulation and Risk Assessment of Weather-Related Cascading Outages}
%



\maketitle

\begin{abstract}
Weather and environmental factors are verified to have played significant roles in historical major cascading outages and blackouts. Therefore, in the simulation and risk assessment of cascading outages in power systems, it is necessary to consider the weather and environmental effects. This paper proposes a method for the risk assessment of weather-related cascading outages. Based on the analysis of historical outage records and temperature-dependent physical outage mechanisms of transmission lines, an outage rate model considering weather condition and conductor temperature is proposed, and the analytical form of outage probability of lines are derived. With the weather-dependent outage model, a two-stage risk assessment method based on Markovian tree (MT) search is proposed, which consists of offline full assessment, and online efficient update of risk assessment results and continued MT search using updated NWP data. The test cases on NPCC 140-bus test system model in winter and summer scenarios verify the advantages of the proposed risk assessment method in both accuracy and efficiency. 
\end{abstract}


%
\IEEEpeerreviewmaketitle

\section{Introduction}
\IEEEPARstart{W}{eather} and environmental factors have ineligible effects on the reliability of power systems\cite{yao2016risk}. Since power systems cover vast geographical areas and large numbers of components are directly exposed to the environment, adverse weather conditions deteriorate the working conditions of the components and may even directly cause components to quit. Weather have been verified as a major cause and contributing factor of outages of elements in power systems\cite{kezunovic2008impact}. Many historical cascading outages and blackouts were also attributed to the adverse weather conditions, including hot and low-wind weather \cite{us2004final, liu1998discussion}, thunder storm \cite{berizzi2004italian}, ice and snow \cite{zhou2011great}, etc. Therefore, it is necessary to consider weather conditions in the risk assessment of cascading outages.

The risk assessment of cascading outages considering weather conditions in real applications has two requirements: 1) accurate forecast of weather conditions covering the studied power system; 2) a weather-dependent model of cascading outages. The first requirement is now generally satisfied by the numerical weather prediction (NWP) service \cite{hosek2011effect}, whose results can be retrieved from public resources. The weather-dependent model for cascading outage simulation that can be utilized along with the NWP has not been proposed. Ref. \cite{rios2002value} proposed a method for simulation of weather-related cascading outages, but the method does not provide time elapse of cascading outages, which is hard to be utilized along the weather data time series. Also it still uses Monte-Carlo method to assess the risk, which is inefficient and difficult to use in online applications. Weather conditions are more extensively considered in the reliability assessment of power systems \cite{billinton2006application,wang2002reliability,bollen2000effects}. However, these approaches are mainly for long-term reliability assessment, not for the online analysis of dependent cascading outages. Also, the typical weather models in these approaches (e.g. the two-state/three-state model) are too coarse to meet the requirement of operational risk assessment. 

During the evolution process of cascading outages, the weather type, quantitative weather condition, and the working conditions influenced by the environment (e.g. the temperature of transmission line conductors) have significant influence on the outage rate of the elements. Therefore, a more accurate weather-dependent element outage model is necessary. Ref. \cite{yang2016interval} proposed an interval method for the estimation of outage rates of transmission lines on the condition of given weather scenarios, which meets the requirements for the risk assessment of 1-hour resolution. Moreover, to meet the requirement for the risk assessment of cascading outages, the outage rate as a function of real-time states should also be considered. The record of Aug. 14 2003 blackout \cite{us2004final} indicates that the dynamic of conductor temperature variation should be analyzed in timescale of a sub-hourly level. Our previous work \cite{yao2017efficient} has proposed an efficient method for the simulation of transmission line temperature evolution (TLTE) based on NWP data. Also, \cite{yao2016risk} proposed an efficient method for the risk assessment of multi-timescale cascading outages \cite{7254205} based on Markovian tree (MT) search. The method provides time elapse in cascading outage processes, so that it can be utilized for the risk assessment of weather-related cascading outages by using NWP data. 

This paper proposes a method for the risk assessment of weather-related cascading outages. The paper first proposes a generic model of transmission line outage depending on the weather condition and conductor temperature, and the analytical expression of the outage probability is derived from the generic outage model, which sets foundation for the efficient risk assessment considering environmental factors. Also from the major mechanisms of line outages, the detailed method for determining the parameters in the outage model is provided. With the weather-dependent outage model, a two-stage method for cascading outage risk assessment based on Markovian tree (MT) search is proposed. The method consists of offline full assessment using the possibly less accurate weather data, and online efficient update of risk assessment result as well as continued MT search with latest weather data, which enhances both accuracy and efficiency as compared to the single-stage offline or online assessment.

The rest of the paper is organized as follows. Section II proposes the generic model of transmission line outage rate, and derives the analytical form of outage probability. Section III elucidates the determination of the transmission line outage rates based on weather data, historical data and conductor specifications. Section IV introduces the two-stage weather-dependent risk assessment of cascading outages. Section V is the test case on the risk assessment of NPCC system model in winter and summer scenarios. Section VI is the conclusion.

\section{Weather-dependent transmission line outage model}
%
%
%
%

\subsection{Outage rates of transmission lines}
In order to model transmission line outages, we need to analyze the in-depth mechanism of line outages. Some causes of line outages have strong correlation with excessive conductor temperature, e.g. tree contacts due to sagging and line outage due to annealing \cite{wan1999increasing}. While some other causes do not have strong relationship with conductor temperature, but highly depend on the weather conditions. Therefore line outages should be described as stochastic processes whose parameters are dependent on weather conditions and conductor temperature. Assume the occurrence of line outage follows Poisson process, then its parameter, the failure rate $\lambda$ is a function of conductor temperature $T_c$ and weather condition $\mathbf{e}$.
Because of the routine utility vegetation management (UVM) along the transmission line and the factory temperature rating of conductor materials, the transmission line below a certain temperature $T_{safe}$ (e.g. $70^\circ\mathrm{C}$) is regarded as free from outages caused by over-temperature of conductor. In this case the outage rate is assumed as a function of weather condition $\mathbf{e}$. 

\begin{equation}\label{eqn:outage_rate_under_safe}
\lambda(T_c,\mathbf{e}|T_c<T_{safe})=\lambda_s(\mathbf{e})
\end{equation}

The historical records of line outages can roughly distinguish the cause and the environmental condition at the occurrence of an outage. Given a certain weather condition $\mathbf{e}$, the overall conditional outage rate of the studied line $\overline{\lambda}(\mathbf{e})$ can be estimated. Further considering the factor of conductor temperature $T_c$, $\overline{\lambda}(\mathbf{e})$ can be expressed as

\begin{equation}\label{eqn:avg_outage_rate}
\begin{aligned}
& \overline{\lambda}(\mathbf{e})=\int_{-\infty}^{+\infty}{\lambda(T_c,\mathbf{e})p_T(T_c)}\mathrm{d}T_c \\
& =\int_{-\infty}^{T_{safe}}{\lambda(T_c,\mathbf{e})p_T(T_c)}\mathrm{d}T_c+
\int_{T_{safe}}^{+\infty}{\lambda(T_c,\mathbf{e})p_T(T_c)}\mathrm{d}T_c\\
& =\int_{-\infty}^{T_{safe}}{\lambda_s(\mathbf{e})p_T(T_c)}\mathrm{d}T_c+
\int_{T_{safe}}^{+\infty}{\lambda(T_c,\mathbf{e})p_T(T_c)}\mathrm{d}T_c
\end{aligned}
\end{equation}

\noindent where $p_T(\cdot)$ is the probabilistic distribution function of $T_c$. Since the transmission line outages caused by overheating and excessive temperature only accounts for a very small portion over all the outages \cite{BPA10years}, the second term in (\ref{eqn:avg_outage_rate}) is much smaller than the first term, so approximately

\begin{equation}\label{eqn:avg_outage_rate_approx}
\overline{\lambda}(\mathbf{e})\approx \int_{-\infty}^{T_{safe}}{\lambda_s(\mathbf{e})p_T(T_c)}\mathrm{d}T_c=\lambda_s(\mathbf{e})Pr(T_c<T_{safe})
\end{equation}

Since the overall probability of transmission line temperature exceeding safe level is rare,  $Pr(T_c<T_{safe})\approx 1$, so

\begin{equation}\label{eqn:avg_outage_rate_approx1}
\lambda_s(\mathbf{e})\approx \overline{\lambda}(\mathbf{e})
\end{equation}

Below $T_{safe}$ the outage rate can be approximated as statistical failure rate $\overline{\lambda}(\mathbf{e})$, while above $T_{safe}$ additional risk of outage emerges, e.g. tree contact due to sagging, and conductor damage due to annealing, etc. So the outage rate above $T_{safe}$ will be larger than $\overline{\lambda}(\mathbf{e})$ and depends on the temperature of conductor $T_c$, so the outage rate $\lambda$ can be regarded the sum of $\overline{\lambda}(\mathbf{e})$ and additional outage rate $\lambda_c(T_c,\mathbf{e})$ .

\begin{equation}\label{eqn:outage_rate_component}
\lambda(T_c,\mathbf{e})=\overline{\lambda}(\mathbf{e})+\lambda_c(T_c,\mathbf{e})
\end{equation}

\begin{figure}[htb]
	\centering
	\includegraphics[width=4cm]{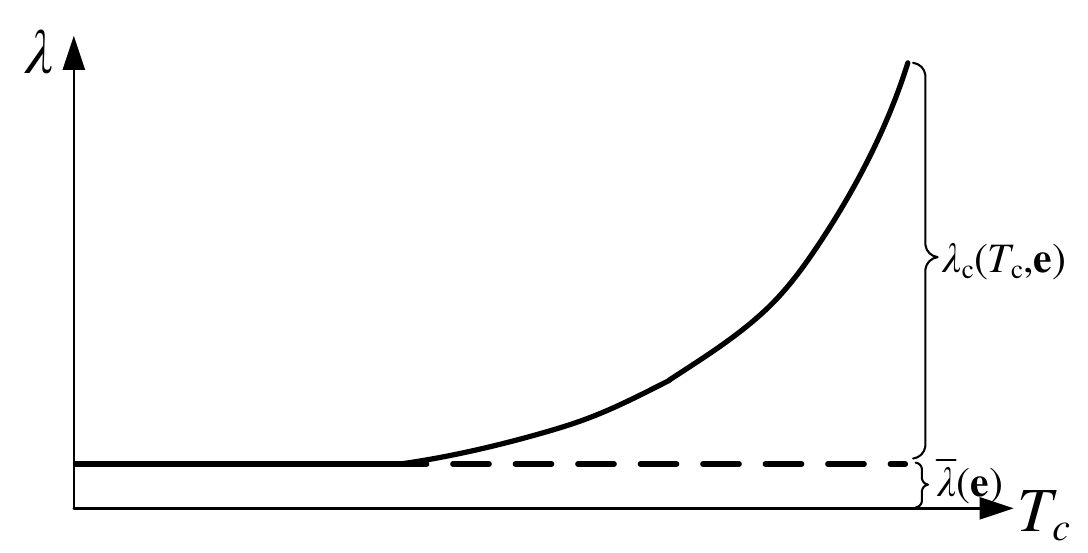}\\
	\caption{Failure rate as a function of conductor temperature}\label{fig:lambda_temperature}
\end{figure}

\subsection{Analytical weather-dependent line outage rate}
Our previous work \cite{yao2017efficient} proposes an approximate analytical solution to the transmission line temperature evolution (TLTE). The method periodically retrieves the environmental parameters such as ambient temperature, wind speed and direction, etc., from numerical weather prediction (NWP), and then derives the parameters of analytical solution from the environmental parameters, conductor parameters and line current. The approximate solution is in the following form
\begin{equation}\label{eqn:temp_ana_solution}
T_{ci}(t)=T_{ei}(p_i(t_0))+(T_{ci}(t_0)-T_{ei}(p_i(t_0)))\mathrm{e}^{-\beta_i(p_i(t_0))(t-t_0)}
\end{equation}

\noindent where $T_{ci}(t)$ is the temperature of line segment $i$ at time $t$. $p_i$ are all the factors that influence the TLTE, such as
weather, line current, conductor parameters, etc. The NWP result is collected periodically. Currently the temporal resolution of NWP covering the United States is 15 min, which can satisfy the requirement of TLTE analysis \cite{hosek2011effect}. Due to the thermal volume of the line conductor, the TLTE is not sensitive to the fast fluctuation of the parameters. So a set of constant values $p_i(t_0)$ is used to represent the average effect of the parameters over the 15min period. $T_{ei}$ and $\beta_i$ are the parameters of the analytical solution, which depend on $p_i$. Within the 18 hours time coverage of NWP, the TLTE of each line segment can be expressed as joined segments of 15-min analytical solutions. 

Assume that the unit-length failure rate of the line segment $\lambda_i(T_{ci})$ is a continuous function of conductor temperature in a certain range of $[\underline{T_{ci}},\overline{T_{ci}}]$, and then it can be approximated with a polynomial of $T_{ci}$.

\begin{equation}\label{eqn:lambda_poly}
\lambda_i(T_{ci}) = \sum_{k=0}^{n}a_{ik} {T_{ci}}^k, T_{ci}\in [\underline{T_{ci}},\overline{T_{ci}}]
\end{equation}


After obtaining (\ref{eqn:lambda_poly}), substituting (\ref{eqn:temp_ana_solution}) into (\ref{eqn:lambda_poly}) gets
\begin{equation}\label{eqn:lambda_time}
\begin{aligned}
&\lambda_{it}(t)=\lambda_i(T_{ci}(t))\\
&=\sum_{k=0}^{n}a_{ik} {\left(T_{ei}+(T_{ci}(t_0)-T_{ei})\mathrm{e}^{-\beta_i(t-t_0)}\right)}^k\\
&=\sum_{k=0}^{n}b_{ik} \mathrm{e}^{-k\beta_i(t-t_0)}
\end{aligned}
\end{equation}

\noindent where the coefficient $b_{ik}$ are
\begin{equation}\label{key}
b_{ik}=\sum_{j=k}^{n} a_{ij} \frac{j!}{k!(j-k)!} {T_{ei}}^{j-k} (T_{ci}(t_0)-T_{ei})^k
\end{equation}

The integral of failure rate over time is 

\begin{equation}\label{eqn:integral_lambda}
\int_{t_0}^{t}\lambda_{it}(t)\mathrm{d}t=b_{i0}(t-t_0)+\sum_{k=1}^{n}\frac{b_{ik}}{k\beta_i}-\sum_{k=1}^{n}\frac{b_{ik}}{k\beta_i}\mathrm{e}^{-k\beta_i(t-t_0)}
\end{equation}

The overall failure rate of a transmission line $L_j$ is the sum of failure rates of all line segments belonging to $L_j$.
\begin{equation}\label{eqn:failure_rate_line}
\lambda_{L_jt}(t)=\sum_{i\in L_j}l_i\cdot \lambda_{it}(t)
\end{equation}

\noindent where $l_i$ is the length of segment $i$.
During the time interval of $[t_0,t]$, the probability of line $L_j$ failure is
\begin{equation}\label{eqn:line_failure_prob}
\mathrm{Pr}_{L_j}(t)=1-\mathrm{e}^{-\sum_{i\in L_j}l_i\cdot\int_{t_0}^{t}\lambda_{it}(t)\mathrm{d}t}
\end{equation}

The Markovian tree model uses the probability of line $L_j$ first trip during the interval $[t_0,t_0+\tau_D]$, which is
\begin{equation}\label{eqn:line_failure_prob_mt}
\mathrm{Pr}_{L_j}^{\mathrm{MT}}(t_0+\tau_D)=\int_{t_0}^{t_0+\tau_D}\lambda_{L_jt}(t)
\mathrm{e}^{-\sum_{L_j}\int_{t_0}^{t}\lambda_{L_jt}(\tau)\mathrm{d}\tau}\mathrm{d}t
\end{equation}

For general $\lambda_{L_jt}(t)$, (\ref{eqn:line_failure_prob_mt}) does not have an explicit analytical form, but it can be approximately calculated by converting integral into a numerical summation. If $\lambda_{L_jt}(t)$ is not time-dependent, then (\ref{eqn:line_failure_prob_mt}) has analytical form

\begin{equation}\label{eqn:line_failure_prob_mt_notime}
\mathrm{Pr}_{L_j}^{\mathrm{MT}}(t)=\frac{\lambda_{L_jt}}{\sum_{L_j}\lambda_{L_jt}}
\left( 1- \mathrm{e}^{\sum_{L_j}\lambda_{L_jt}(t-t_0)} \right)
\end{equation}

Remark: the method does not require that the failure rate is a continuous function of $T_{ci}$. It is, rather assumed as a non-decreasing function of $T_{ci}$. Since the temperature evolution (\ref{eqn:temp_ana_solution}) is monotonic, if there is discontinuity, then $\lambda_i(T_{ci})$ can be treated in segments.

\section{Determination of outage rates}
This section proposes a detailed method for determining the outage rates of transmission lines. According to (\ref{eqn:outage_rate_component}) in Section II.A, the outage rate can be decomposed as the sum of average weather-dependent outage rate $\overline{\lambda}(\mathbf{e})$ and additional outage rate of high conductor temperature $\lambda_c(T_c,\mathbf{e})$. We assume the threat of line outage at high conductor temperature is excessive sagging and damage caused by annealing, so $\lambda_c(T_c,\mathbf{e})$ is regarded as the sum of outage rates caused by sagging $\lambda_{Sag}(T_c,\mathbf{e})$ and $\lambda_{Anneal}(T_c,\mathbf{e})$. Then the overall outage rate is 

\begin{equation}\label{eqn:outage_rate_3}
\lambda(T_c,\mathbf{e})=\overline{\lambda}(\mathbf{e})+\lambda_{Sag}(T_c,\mathbf{e})+\lambda_{Anneal}(T_c,\mathbf{e})
\end{equation}

The following subsections elucidate the determination of every component of outage rate.

\subsection{Failure rate related to different weather phenomena}
Weather conditions can significantly influence the reliability of elements in power systems. Take transmission lines as an example. Their operation can be affected by various adverse weather conditions, as listed in Table. I.

\begin{table}[htb]
	\centering
	\caption{Hazards of weather conditions to transmission lines}
	\label{tab:hazard_weather}
	\begin{tabularx}{\linewidth}{p{3.4cm}X}
		\toprule[1.5pt]
		\textbf{Weather condition} $\mathbf{e}$  & \textbf{Cause of outage} \\\midrule[1pt]
		rain/fog/haze               & fog-/wet-flashover\\
		thunderstorm                & lightning strike\\
		high wind                   & increased tension \& tree contact\\
		snow/freezing rain          & increased tension load\\
		hot weather                 & increased conductor temperature\\
		\bottomrule[1.5pt]
	\end{tabularx}
\end{table}

\cite{yang2016interval} proposed a method for inferring the weather-related outage rates $\overline{\lambda}(\mathbf{e})$ from historical records. Therefore, with the forecast of weather conditions from numerical weather prediction, the outage rates of transmission lines in the system over a period can be estimated. In this paper, we use the method in \cite{yang2016interval} to estimate $\overline{\lambda}(\mathbf{e})$ from historical records.


\subsection{Failure rate due to sagging and tree contact}

The sag $s$ of line section can be approximated as a linear function of conductor temperature $T_c$ \cite{qi2013blackout}.
\begin{equation}\label{eqn:sag_to_temperature}
s(T_c)=s_0+s_T\cdot T_c
\end{equation}

The higher $T_c$, the larger $s(T_c)$, and the higher is the chance that trees or objects intrude into the discharging range of conductors and cause line outage. Define the sag that just causes line outage as critical sag $s_c$, and assume the distribution of critical sag (determined mainly by object heights) to satisfy normal distribution, so the temperature corresponding to $s_c$ (named critical sag temperature  $T_{cs}$)also satisfies normal distribution, as $N(\mu_T,{\sigma_T}^2)$. Assume the rate of failure due to sagging $\lambda_{Sag}$ to be proportional to the probability of conductor temperature $T_c$ being higher than $T_{cs}$.

\begin{equation}\label{eqn:sag_rate}
\lambda_{Sag}(T_c) \propto Pr\{T_{cs}\leq T_c \}=F_{T_{cs}}(T_c)
\end{equation}

\noindent where $F_{T_{cs}}(\cdot)$ is the CDF of $T_{cs}$. Note that due to UVM scheme, usually the vegetation will be trimmed so that under normal operation temperature the line outage due to tree contact is practically avoided. Assume that the post-UVM distribution of $T_{sc}$ satisfies $T_{safe}$ is at least $3\sigma_T$ away from $\mu_T$, which means that tree-contact under $T_{safe}$ is nearly impossible. But due to some extreme hot and low-wind weather, the conductor temperature may drift above $T_{safe}$, causing tree-contact more possible \cite{qi2013blackout, us2004final}.

\subsection{Failure rate due to annealing and tensile loss}


The tensile loss characteristics under elevated temperature of various conductors are proposed in \cite{harvey1972effect,morgan1996effect,morgan1979loss}. For example, 
\cite{harvey1972effect} proposed the formula on percentage of tensile loss of aluminum conductors.
\begin{equation}\label{eqn:loss_tensile}
La=100-T\cdot t^{-0.00254\max\{ T_c-95,0 \}/D}
\end{equation}

\noindent where $T=\min\{ 134-0.24T_c,100 \}$ in ${}^\circ \mathrm{C}$, conductor diameter $D$ is in mm, and the duration $t$ is in hour. Assume the outage rate due to annealing is proportional to the loss of tensile

\begin{equation}\label{eqn:anneal_rate}
\lambda_{Anneal}(T_c)\propto La
\end{equation}

\begin{figure}[htb]
	\centering
	\includegraphics[width=4cm]{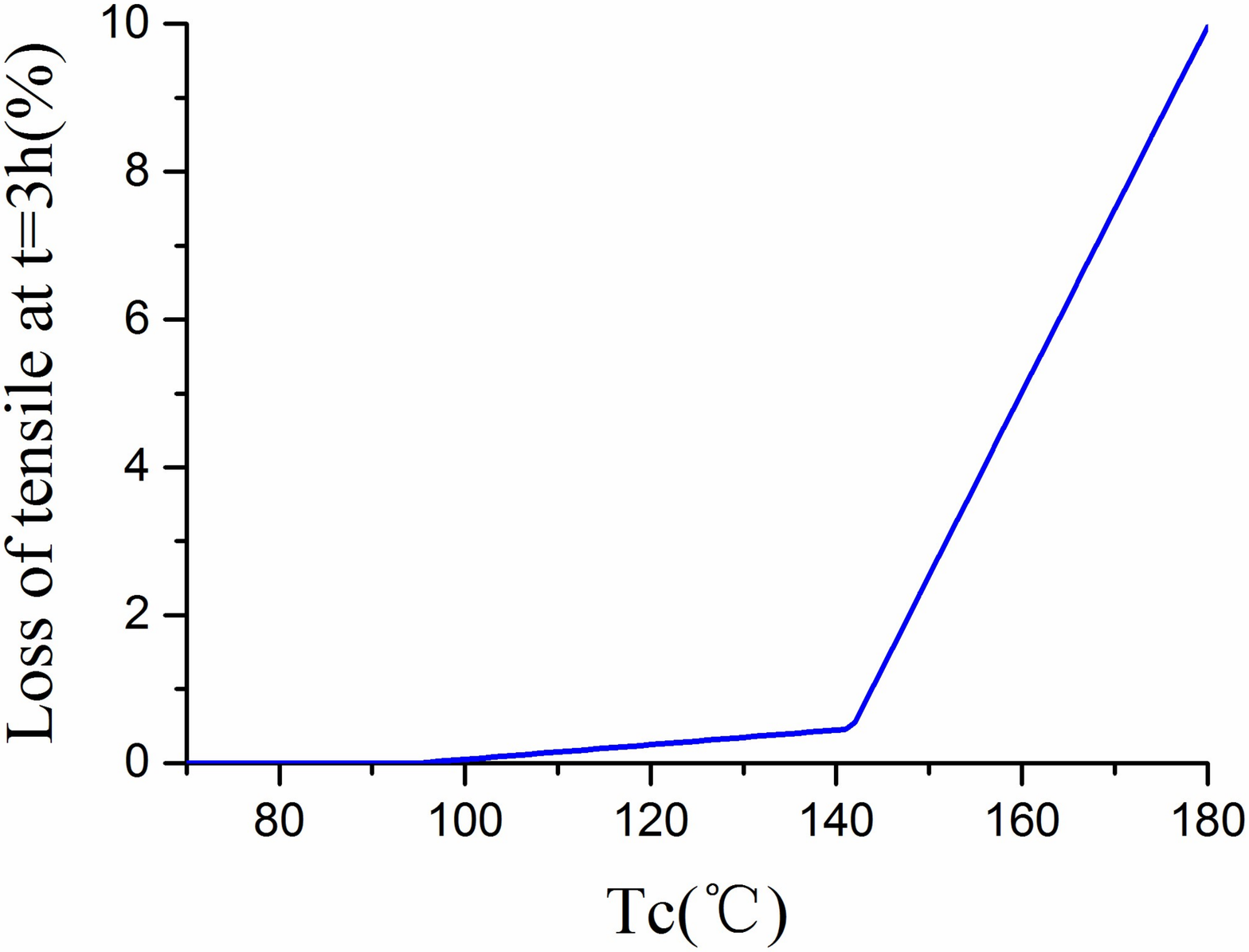}\\
	\caption{Loss of tensile as function of conductor temperature}\label{fig:tensile_loss}
\end{figure}

(\ref{eqn:loss_tensile}) indicates that the tensile starts to decrease upon reaching $95^\circ\mathrm{C}$, and loss of tensile significantly accelerates after temperature becomes higher than around $140^\circ\mathrm{C}$, as shown in Fig. \ref{fig:tensile_loss}.

\section{Two-stage risk assessment with MT search}
Our previous work \cite{yao2016risk} proposed an efficient risk assessment of cascading outages method based on Markovian tree search. The method derives an expansion of cascading outage risk corresponding to the tree structure. Here we assume that the initial power flow pattern for cascading outage analysis does not change substantially from the day-ahead unit-commitment result to the real-time power flow, which is generally true in most cases. In this case, when the environmental condition changes or is updated, the terms of the outage probabilities in the expansion can be re-calculated and thus the total risk is efficiently updated. This scheme avoids the re-calculation of entire risk assessment. 

Therefore, a two-stage scheme of risk assessment based on Markovian tree search is proposed. The offline stage is to conduct complete risk assessment with the available weather data, which has no big difference from the original version proposed in \cite{yao2016risk}. Then the online stage refines the result of risk assessment with updated weather data, updates the risk estimation index on MT, and continues several attempts of MT search to find new risk brought by the updated weather data. The online stage includes the following two steps.

\subsection{Update the risk terms on MT}

According to \cite{yao2016risk}, the risk assessment based on Markovian tree search derives the expansion of cascading outage risk corresponding to the tree structure:

\begin{equation}\label{eqn:MT_expansion}
\begin{aligned}
& R=C_0+\sum_{k_1}\mathrm{Pr}(i_{k_1})C(i_{k_1})\\
& +\sum_{k_1}\mathrm{Pr}(i_{k_1})\sum_{k_2}\mathrm{Pr}(i_{k_2}|i_{k_1})C(i_{k_1}i_{k_2}) \\
& +\sum_{k_1}\mathrm{Pr}(i_{k_1})\sum_{k_2}\mathrm{Pr}(i_{k_2}|i_{k_1})\sum_{k_3}\mathrm{Pr}(i_{k_3}|i_{k_1}i_{k_2})C(i_{k_1}i_{k_2}i_{k_3})\\
&+\cdots
\end{aligned}
\end{equation}

\noindent each level on the MT corresponds to a time interval $\tau_D$, and each node is labeled with the outage sequence from the root, as $(i_{k_1}\cdots i_{k_n})$, where $i_{k_r}$ is either a positive integer denoting the index of the element failed on the $r$th level or 0 if no outage occurs on this level. The cost corresponding to state $(i_{k_1},\cdots,i_{k_n})$ is $C(i_{k_1} \cdots i_{k_n})$, and the conditional probability of outage event $i_{k_{r+1}}$ after state $(i_{k_1} \cdots,i_{k_r})$ is $\mathrm{Pr}(i_{k_{r+1}}|i_{k_1} \cdots i_{k_r})$.

The update of risk assessment result has two steps:

1) The update of transmission line temperature evolution (TLTE), which is necessary for estimating the outage probability. The update of TLTE is sequential along a cascading outage path. With the analytical solution of TLTE adaptable to multiple system states proposed in \cite{yao2017efficient}, the TLTE can be calculated efficiently.

2) With the weather data and TLTE, the probability terms in the MT expansion of risk is updated according to Section II.B. This step can be implemented efficiently in parallel. 

\subsection{Continue MT search}
The continued MT search requires the update of not only all the risk terms in (\ref{eqn:MT_expansion}), but also the risk estimation indices (REI) that effectively guides the search to high-risk cascading outage paths. So in this step, the REIs should be first updated, and then MT search can be continued.

1) The REIs are updated recursively on the MT. The complete procedure is demonstrated in Algorithm 1.

\begin{table}[htb]
	\centering
	\label{tab:alg1}
	\begin{tabularx}{\linewidth}{X}
		\toprule[1.5pt]
		\textbf{Algorithm 1.} Updating REIs (pseudo-codes). \\\midrule[1pt]
		Assume $Pr^0(\cdot)$ and $\rho^0(\cdot)$ are original outage probability and REI, and updated outage probability and REI are denoted as $Pr(\cdot)$ and $\rho(\cdot)$.\\\midrule[1pt]
		\textbf{function} \textit{updateREI}($i_{k_1}\cdots i_{k_r}$)\\
		\hspace{1em}\textbf{if} $i_{k_1}\cdots i_{k_r}$ is the end of cascade path\\
		\hspace{2em}$\rho_{i_{k_1}\cdots i_{k_r}}=\varepsilon_R$\\
		\hspace{1em}\textbf{elseif} $i_{k_1}\cdots i_{k_r}$ is not reached on MT\\
		\hspace{2em}$\rho_{i_{k_1}\cdots i_{k_r}}=\rho^0_{i_{k_1}\cdots i_{k_r}}\cdot Pr(i_{k_r}|i_{k_1}\cdots i_{k_{r-1}})/ Pr^0(i_{k_r}|i_{k_1}\cdots i_{k_{r-1}})$\\
		\hspace{1em}\textbf{else} (i.e. $i_{k_1}\cdots i_{k_r}$ has subordinate states)\\
		\hspace{2em}\textbf{foreach} subordinate state $i_{k_1}\cdots i_{k_r}i_{k_{r+1}}$\\
		\hspace{3em}\textit{updateREI}($i_{k_1}\cdots i_{k_r}i_{k_{r+1}}$)\\
		\hspace{2em}\textbf{endforeach}\\
		\hspace{2em}update probability for MT search\cite{yao2016risk}:\\ 
		\hspace{2em}$Pr_{i_{k_1}\cdots i_{k_r}i_{k_{r+1}}}^{\mathrm{calc}}=\frac{(\rho_{i_{k_1}\cdots i_{k_r}i_{k_{r+1}}})^\mu}{\sum_{i'_{k_{r+1}}}(\rho_{i_{k_1}\cdots i_{k_r}i'_{k_{r+1}}})^\mu}$ ($\mu\geq 0$)\\
		\hspace{2em}update REI: $\rho_{i_{k_1}\cdots i_{k_r}}= \sum_{i'_{k_{r+1}}}\rho_{i_{k_1}\cdots i_{k_r}i'_{k_{r+1}}}Pr_{i_{k_1}\cdots i_{k_r}i'_{k_{r+1}}}^{\mathrm{calc}}$\\
		\hspace{1em}\textbf{endif}\\
		\textbf{end}\\
		\bottomrule[1.5pt]
	\end{tabularx}
\end{table}

2) Continue the risk assessment of MT search\cite{yao2016risk} using the new weather data.

\section{Tests on NPCC 140-bus system}
\subsection{Weather-dependent parameters of outage rates}
The base weather-dependent outage rate is determined using the method proposed in \cite{yang2016interval}. Here the precipitation type is categorized as four: rain, snow, freezing rain and ice pellets. The weather condition evidence variables in this paper are ambient temperature ($E_1$), wind speed ($E_2$), lightning ($E_3$), rain ($E_4$), snow ($E_5$), freezing rain ($E_6$), ice pellets ($E_7$). The evidence and outcome are defined in Table \ref{tab:outage_parameter_evidence}.

\begin{table}[htb]
	\centering
	\caption{Evidence and outcome variables}
	\label{tab:outage_parameter_evidence}
	\begin{tabularx}{\linewidth}{p{0.8cm}ccc}
		\toprule[1.5pt]
		Variable  & State 1      & State 2     & State 3 \\ \midrule[1pt]
		\centering\multirow{2}*{$E_1$}  & $T_a<4^\circ\mathrm{C}$ & $4^\circ\mathrm{C}\leq T_a< 26^\circ\mathrm{C}$ & $T_a\geq 26^\circ\mathrm{C}$   \\
		 & ($e_{1,1}$) & ($e_{1,2}$) & ($e_{1,2}$) \\
		\centering\multirow{2}*{$E_2$} & $V_w<12$km/h & $12$km/h$\leq V_w<40$km/h & $V_w\geq 40$km/h\\
		 & ($e_{2,1}$) & ($e_{2,2}$) & ($e_{2,2}$) \\
		 \centering\multirow{2}*{$E_3$} & Lightning & No lightning & -- \\
		 & ($e_{3,1}$) & ($e_{3,2}$) & --  \\
		 \centering\multirow{2}*{$E_4$} & Rain & No rain & -- \\
		 & ($e_{4,1}$) & ($e_{4,2}$) & --  \\
		 \centering\multirow{2}*{$E_5$} & Snow & No snow & -- \\
		 & ($e_{5,1}$) & ($e_{5,2}$) & --  \\
		 \centering\multirow{2}*{$E_6$} & Freezing rain & No freezing rain & -- \\
		 & ($e_{6,1}$) & ($e_{6,2}$) & --  \\
		 \centering\multirow{2}*{$E_7$} & Ice pellets & No ice pellets & -- \\
		 & ($e_{7,1}$) & ($e_{7,2}$) & --  \\
		\bottomrule[1.5pt]
	\end{tabularx}
\end{table}

Assume $h_1$ stands for the normal state of a transmission line, and $h_2$ stands for outage. The outage rate parameters are the same as in Sections V.A and V.B in \cite{yang2016interval} besides the parameters for freezing rain and ice pellets, as shown in Table \ref{tab:outage_parameter_e6} and \ref{tab:outage_parameter_e7}.

\begin{table}[htb]
	\centering
	\caption{Conditional mass function of freezing rain}
	\label{tab:outage_parameter_e6}
	\begin{tabularx}{0.75\linewidth}{p{0.8cm}ccc}
		\toprule[1.5pt]
		$H$    & $E_1$     & $P(e_{6,1}|H,E_1)$ & $P(e_{6,2}|H,E_1)$ \\ \midrule[1pt]
		$h_1$  & $e_{1,1}$ & 0.14               & 0.86               \\
		$h_1$  & $e_{1,2}$ & 0.03               & 0.97               \\
		$h_1$  & $e_{1,3}$ & 0.00               & 1.00               \\
		$h_2$  & $e_{1,1}$ & [0.85, 0.95]       & [0.05, 0.15]      \\
		$h_2$  & $e_{1,2}$ & [0.00, 0.03]       & [0.97, 1.00]      \\
		$h_2$  & $e_{1,3}$ & [0.00, 0.01]       & [0.99, 1.00]      \\
		\bottomrule[1.5pt]
	\end{tabularx}
\end{table}


\begin{table}[htb]
	\centering
	\caption{Conditional mass function of ice pellets}
	\label{tab:outage_parameter_e7}
	\begin{tabularx}{0.75\linewidth}{p{0.8cm}ccc}
		\toprule[1.5pt]
		$H$    & $E_1$     & $P(e_{7,1}|H,E_1)$ & $P(e_{7,2}|H,E_1)$ \\ \midrule[1pt]
		$h_1$  & $e_{1,1}$ & 0.14               & 0.86               \\
		$h_1$  & $e_{1,2}$ & 0.02               & 0.98               \\
		$h_1$  & $e_{1,3}$ & 0.00               & 1.00               \\
		$h_2$  & $e_{1,1}$ & [0.80, 0.90]       & [0.10, 0.20]      \\
		$h_2$  & $e_{1,2}$ & [0.00, 0.02]       & [0.98, 1.00]      \\
		$h_2$  & $e_{1,3}$ & [0.00, 0.01]       & [0.99, 1.00]      \\
		\bottomrule[1.5pt]
	\end{tabularx}
\end{table}

Assume the number of evidence is $n_E$, and the bounds of the outage rate are:
\begin{equation}\label{eqn:outage_rate_lower_bnd}
\begin{aligned}
&\underline{\lambda}=\underline{Pr}(h_2|\mathbf{e})=
\min{}\frac{\prod_{j=1}^{n_E}Pr(e_{j,k_j}|h_2,\mathbf{e})\cdot Pr(h_2)}
{\sum_{i=1}^{2}\prod_{j=1}^{n_E}Pr(e_{j,k_j}|h_i,\mathbf{e})\cdot Pr(h_i)}\\
&\overline{\lambda}=\overline{Pr}(h_2|\mathbf{e})=
\max{}\frac{\prod_{j=1}^{n_E}Pr(e_{j,k_j}|h_2,\mathbf{e})\cdot Pr(h_2)}
{\sum_{i=1}^{2}\prod_{j=1}^{n_E}Pr(e_{j,k_j}|h_i,\mathbf{e})\cdot Pr(h_i)}\\
\end{aligned}
\end{equation}

\noindent where $\mathbf{e}=\{e_{j,k_j}\},j=1,2,\cdots,n_E$. In this case, we use the median value as the estimation of outage rate, $\lambda=(\underline{\lambda}+\overline{\lambda})/2$. 

\subsection{Winter scenario}
In this section, we test the proposed method on the NPCC 140-bus system. 
Select the weather conditions at 12:00 EDT (Eastern Daylight Time) on March 14th, 15th and 16th, 2017, and implement risk assessments respectively. It should be noted that during March 13th-15th, the winter storm "Stella" swept from southwest to northeast through the area of NPCC system. The winter storm reached at maximum in the area on March 14th, and then the storm diminished and nearly completely moved out of NPCC area on 16th. 

\begin{figure}[htb]
	\centering
	\includegraphics[width=6cm]{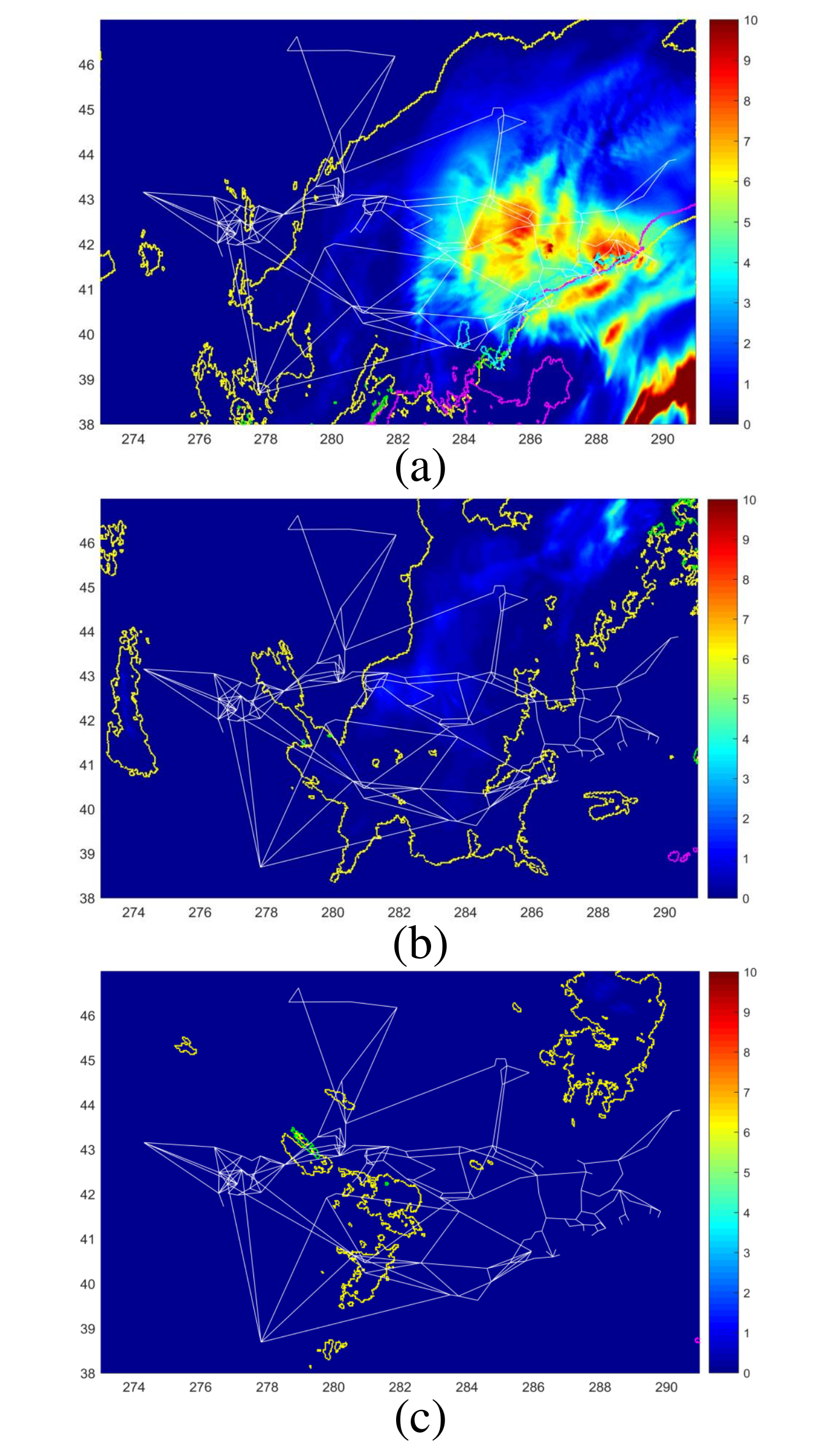}\\
	\caption{1-hour precipitation forecast (mm) at 12:00 EDT on Mar. 14 (a), Mar. 15 (b), and Mar. 16 (c), 2017. Precipitation type by boundary color: magenta -- rain, yellow -- snow, green -- freezing rain, cyan -- ice pellets.}\label{fig:weather_170314}
\end{figure}

Use the weather-dependent risk assessment to calculate risk on these three days, the results are listed in Table \ref{tab:npcc_risk_winter}. The initial system states on the three days are assumed to be the same, and the only difference is in their weather conditions. The result verifies that the weather has significant influence on the cascading outage risk. 

\begin{table}[htb]
	\centering
	\caption{Comparison of cascading outage risk (winter scenario)}
	\label{tab:npcc_risk_winter}
	\begin{tabularx}{\linewidth}{p{0.1cm}@{}*4{>{\centering\arraybackslash}X}@{}}
		\toprule[1.5pt]
		&           & March 14& March 15& March 16\\ \midrule[1pt]
		& Risk(MW)  & 2.48    & 0.54    & 0.11    \\
		\bottomrule[1.5pt]
	\end{tabularx}
\end{table}

First take the cascading outage path with highest risk (path A) on March 14. Table \ref{tab:npcc_risk_winter_path1_rates} lists the outage rates of each level of branch outage. The short-timescale outages are listed in the parenthesis, and the data in bold is the highest outage rate among the three days. It can be seen that the outage rates on March 14 is significantly higher than those on March 15 and 16 for most levels of the cascading outages. Even though the outage rates on some levels are not the highest on March 14, the difference is not significant. 


\begin{table}[htb]
	\centering
	\caption{Outage rates by levels in cascading outage path A}
	\label{tab:npcc_risk_winter_path1_rates}
	\begin{tabularx}{\linewidth}{p{0.4cm}@{}*4{>{\centering\arraybackslash}X}@{}}
		\toprule[1.5pt]
		\centering\multirow{2}*{Level} & \centering\multirow{2}*{Outages} & \multicolumn{3}{c}{Outage rates ($\mathrm{s}^{-1}$)}\\ \cline{3-5}
		&      & March 14 & March 15 & March 16\\ \midrule[1pt]
		1 & 101(initial)& --                   & --       & --      \\
		2 & 47 (48, 52) & $\mathbf{\sci{3.85}{-8}}$ & $\sci{1.60}{-8}$ & $\sci{1.89}{-9}$ \\
		3 & 185         & $\sci{2.78}{-9}$ & $\sci{2.79}{-9}$ & $\mathbf{\sci{3.15}{-9}}$ \\
		4 & 114         & $\sci{2.46}{-9}$ & $\sci{2.36}{-9}$ & $\mathbf{\sci{2.96}{-9}}$ \\
		5 & 178         & $\sci{5.47}{-9}$ & $\sci{5.62}{-9}$ & $\mathbf{\sci{6.89}{-9}}$ \\
		6 & 109         & $\mathbf{\sci{6.22}{-7}}$ & $\sci{9.25}{-8}$ & $\sci{5.66}{-9}$ \\
		7 & 15          & $\mathbf{\sci{3.49}{-8}}$ & $\sci{1.48}{-9}$ & $\sci{1.55}{-9}$ \\
		8 & 66          & $\mathbf{\sci{1.19}{-7}}$ & $\sci{7.99}{-8}$ & $\sci{5.17}{-9}$ \\
		9 & 111         & $\sci{1.83}{-8}$ & $\mathbf{\sci{2.23}{-8}}$ & $\sci{1.45}{-9}$ \\
		10& 9           & $\mathbf{\sci{7.20}{-8}}$ & $\sci{3.49}{-9}$ & $\sci{3.75}{-9}$ \\  \midrule[1pt]
		\multicolumn{2}{c}{Risk of path A(MW)} & $\mathbf{\sci{6.63}{-1}}$ & $\sci{2.76}{-1}$ & $\sci{3.25}{-2}$ \\
		\bottomrule[1.5pt]
	\end{tabularx}
\end{table}

\begin{figure}[htb]
	\centering
	\includegraphics[width=6cm]{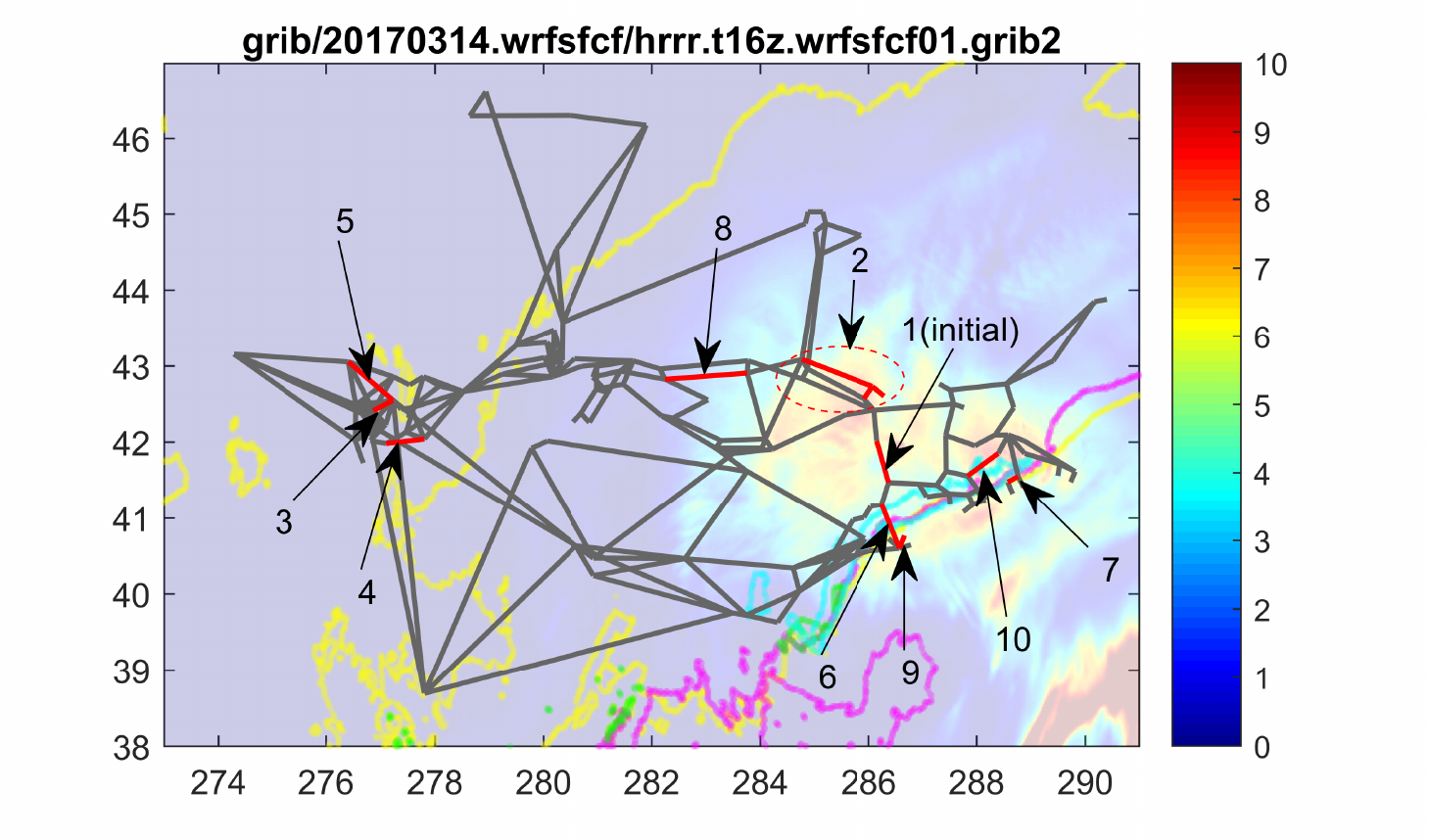}\\
	\caption{Cascading outage path A and weather overlay on March 14.}\label{fig:170314_pathA}
\end{figure}

Fig. \ref{fig:170314_pathA} shows the geographical positions of the lines that are tripped (red) in the cascading outage path A, as well as the weather condition of Mar. 14. The significantly higher outage rates as shown in Table \ref{tab:npcc_risk_winter_path1_rates} correspond to geographical regions with the most severe snow storm. 

\begin{figure}[htb]
	\centering
	\includegraphics[width=6cm]{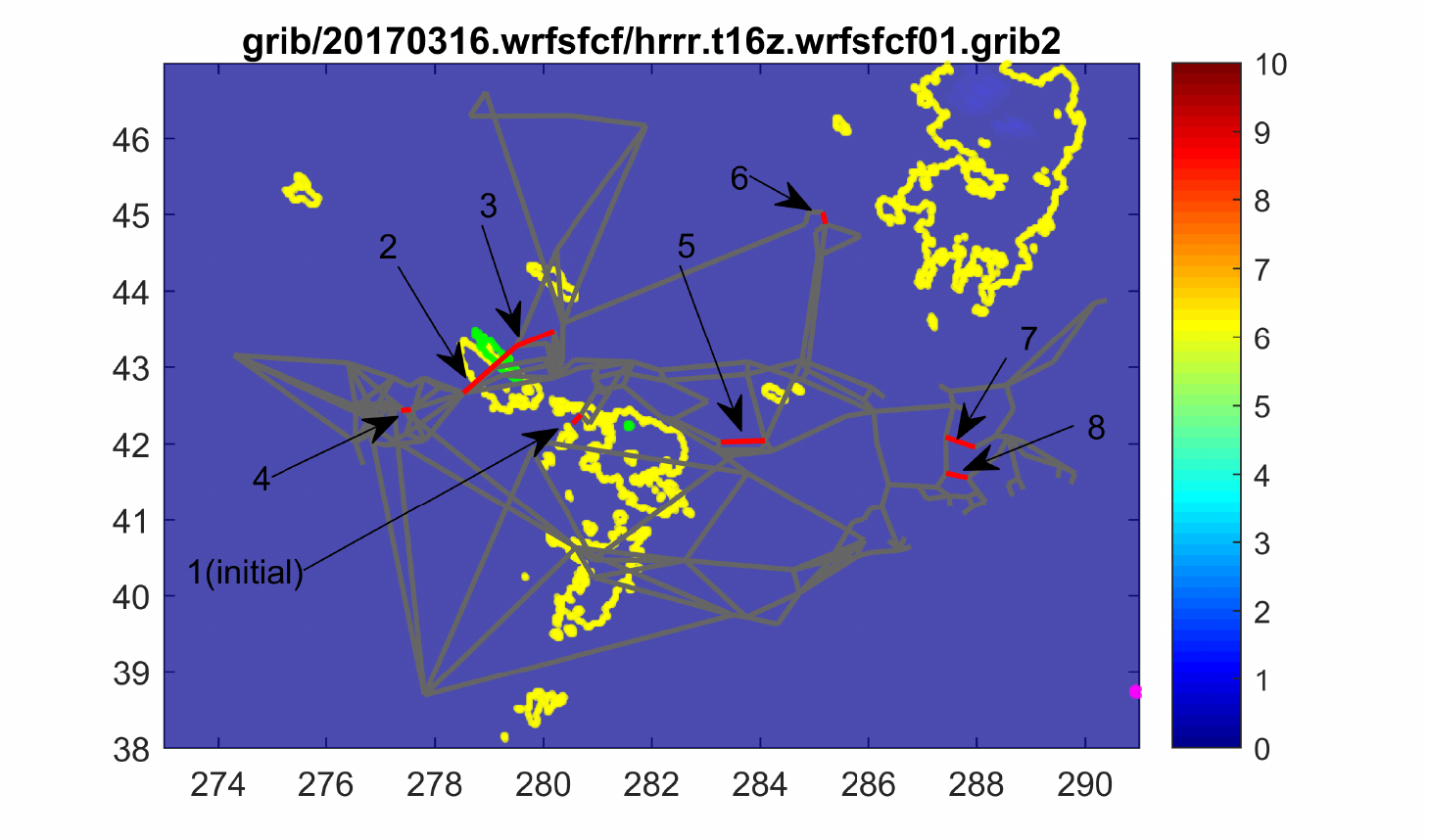}\\
	\caption{Cascading outage path B and weather overlay on March 16.}\label{fig:170316_pathB}
\end{figure}

However, not all the cascading outage paths on March 14 are higher than that on other two days. Take March 16 as an example. Although the overall weather condition on this day is the least risky for cascading outages, some cascading outage paths have higher risk instead. On March 16th, the snow storm has moved out of the NPCC area, but some scattered precipitation still occurred around Lake Erie due to "the lake effect". Correspondingly, the outage rate around the precipitation falling area rises. Table \ref{tab:npcc_risk_winter_path2_rates} lists a cascading outage path (path B, as shown in Fig. \ref{fig:170316_pathB}) that occurs partly around the Lake Erie. 

\begin{table}[htb]
	\centering
	\caption{Outage rates by levels in cascading outage path B}
	\label{tab:npcc_risk_winter_path2_rates}
	\begin{tabularx}{\linewidth}{p{0.4cm}@{}*4{>{\centering\arraybackslash}X}@{}}
		\toprule[1.5pt]
		\centering\multirow{2}*{Level} & \centering\multirow{2}*{Outages} & \multicolumn{3}{c}{Outage rates ($\mathrm{s}^{-1}$)}\\ \cline{3-5}
		&      & March 14 & March 15 & March 16\\ \midrule[1pt]
		1 & 233(initial)& --                   & --       & --      \\
		2 & \textbf{161}& $\sci{5.00}{-8}$ & $\sci{5.63}{-8}$ & $\mathbf{\sci{7.39}{-8}}$ \\
		3 & 162         & $\mathbf{\sci{7.43}{-8}}$ & $\sci{3.77}{-9}$ & $\sci{4.62}{-9}$ \\
		4 & 122         & $\mathbf{\sci{1.39}{-8}}$ & $\sci{1.69}{-9}$ & $\sci{7.38}{-10}$ \\
		5 & 95          & $\sci{9.87}{-8}$ & $\mathbf{\sci{9.89}{-8}}$ & $\sci{4.66}{-9}$ \\
		6 & 58          & $\mathbf{\sci{2.03}{-8}}$ & $\sci{2.02}{-8}$ & $\sci{8.43}{-10}$ \\
		7 & 13          & $\mathbf{\sci{3.07}{-8}}$ & $\sci{1.79}{-9}$ & $\sci{1.89}{-9}$ \\
		8 & 8           & $\mathbf{\sci{5.63}{-8}}$ & $\sci{2.59}{-9}$ & $\sci{2.83}{-9}$ \\  \midrule[1pt]
		\multicolumn{2}{c}{Risk of path B(MW)} & $\sci{9.03}{-5}$ & $\sci{1.02}{-4}$ & $\mathbf{\sci{1.34}{-4}}$ \\
		\bottomrule[1.5pt]
	\end{tabularx}
\end{table}

The results show that the outage rate of line 161 on March 16 is higher than that on March 14 and 15, which is caused by the freezing rain fall along the line. Although rates of other outages are lower on March 16, since the outage of line 161 contributes to most of the risk of the entire cascading outage path, the risk on March 16 is still the highest. 

\subsection{Summer scenario}
\begin{figure}[htb]
	\centering
	\includegraphics[width=8cm]{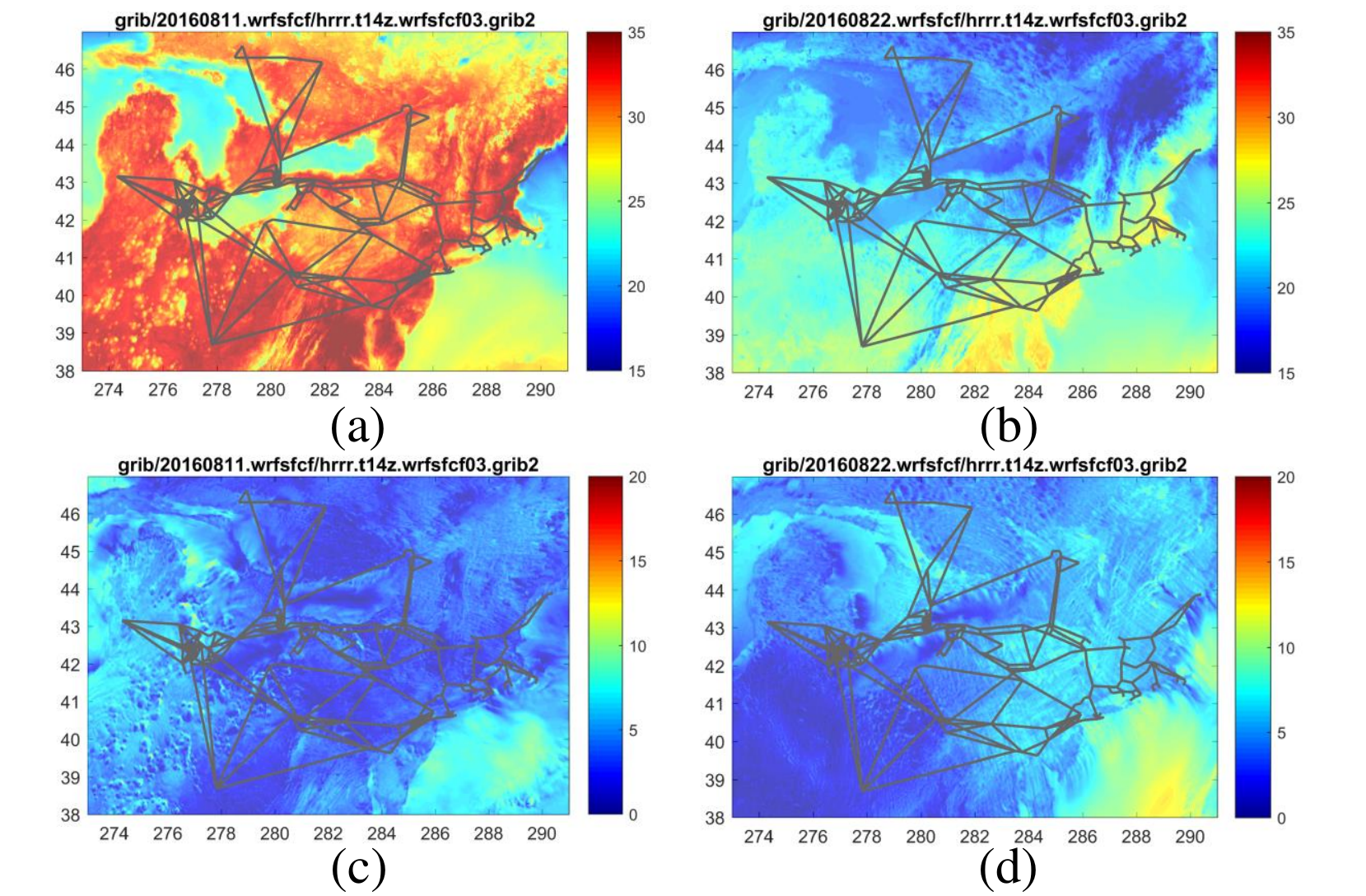}\\
	\caption{Ambient temperature and wind speed at 13:00 EDT, August 11 and 22, 2016.  (a) and (b) are air temperature (${}^{\circ}$C) on these two days, while (c) and (d) show wind speed (m/s).}\label{fig:weather_summer}
\end{figure}

The hottest weather in NPCC area in 2016 occurred during August 11-12. August 11th is a sunny, hot and low-wind day, which may contribute to high conductor temperature of transmission lines. We also choose Aug. 22th, which is sunny, but much cooler and more windy day as comparison. The ambient temperature and wind speed at noon on these two days are demonstrated in Fig. \ref{fig:weather_summer}.

As attributed to the hot and low-wind weather, the cascading outage risk on Aug. 11 is significantly higher (see Table \ref{tab:npcc_risk_summer}).

\begin{table}[htb]
	\centering
	\caption{Comparison of cascading outage risk (summer scenario)}
	\label{tab:npcc_risk_summer}
	\begin{tabularx}{\linewidth}{p{0.1cm}@{}*3{>{\centering\arraybackslash}X}@{}}
		\toprule[1.5pt]
		&           & August 11 & August 22\\ \midrule[1pt]
		& Risk(MW)  & 18.23     & 4.65     \\
		\bottomrule[1.5pt]
	\end{tabularx}
\end{table}

\begin{figure}[htb]
	\centering
	\includegraphics[width=6cm]{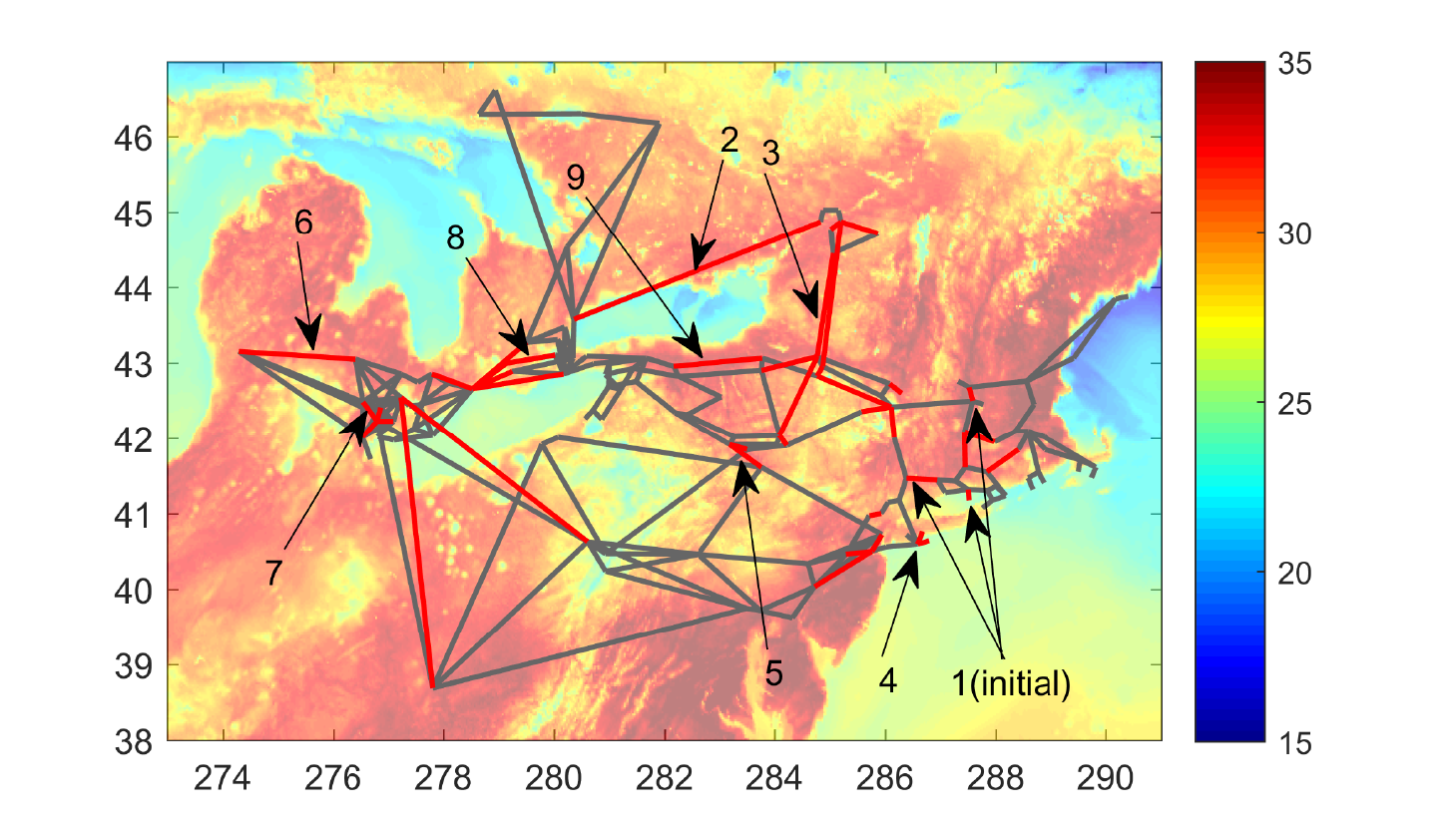}\\
	\caption{Cascading outage path with weather overlay at 13:00 EDT on Aug. 11, 2016.}\label{fig:160811_pathA}
\end{figure}

On Aug. 11, some cascading outage paths with conductor overheat events stand out with much higher risk than on Aug. 22. We take one typical cascading outage path (namely path C) as an example, as shown in Fig. \ref{fig:160811_pathA} and Table. \ref{tab:npcc_risk_summer_path1_rates}. It can be seen that for all outages, the outage rates on August 11 are higher than those on August 22. Such a difference in outage rates contributes to the over 10 times higher risk on Aug. 11 than on Aug. 22. 

\begin{table}[htb]
	\centering
	\caption{Outage rates by levels in cascading outage path C (summer)}
	\label{tab:npcc_risk_summer_path1_rates}
	\begin{tabularx}{\linewidth}{p{2.2cm}@{}*4{>{\centering\arraybackslash}X}@{}}
		\toprule[1.5pt] 
		\centering\multirow{2}*{Outage} & \multicolumn{2}{c}{Aug. 11 } &
		\multicolumn{2}{c}{Aug. 22 } \\ \cline{2-5}
		      & Max $T_c$ (${}^{\circ}$C) & Outage rate (s${}^{-1}$) & 
		Max $T_c$ (${}^{\circ}$C) & Outage rate (s${}^{-1}$) \\ \midrule[1pt]
		 100, 26, 39 (initial) & --                   & --       & --    & --      \\
		 \textbf{141}         & $64.1$  & $\mathbf{\sci{1.81}{-6}}$ & $42.7$ & ${\sci{3.46}{-7}}$ \\
		 \textbf{54}          & $53.4$  & $\mathbf{\sci{3.76}{-7}}$ & $38.1$ & ${\sci{5.47}{-8}}$ \\
		 \textbf{110} (9, 13, 36, 44, 45, 48, 49, 51, 56, 57, 60, 63, 94, 105, 111, 197, 199, 203) 
		        & $72.4$ & $\mathbf{\sci{8.62}{-8}}$ & $42.6$ & ${\sci{2.80}{-8}}$ \\
		 \textbf{88} (96, 128, 149, 151, 153, 161, 224)
		          & $104.4$ & $\mathbf{\sci{9.86}{-7}}$ & $74.8$ & ${\sci{2.32}{-8}}$ \\
		 \textbf{183}         & $102.5$ & $\mathbf{\sci{3.30}{-6}}$ & $76.7$ & ${\sci{1.12}{-7}}$ \\
		 \textbf{190} (191, 192, 193, 216, 228)
		         & $92.3$ & $\mathbf{\sci{5.79}{-7}}$ & $74.5$ & ${\sci{1.72}{-7}}$ \\
		 \textbf{150}         & $83.4$ & $\mathbf{\sci{1.64}{-7}}$ & $69.9$ & ${\sci{6.01}{-8}}$ \\
		 \textbf{65}          & $82.9$ & $\mathbf{\sci{5.25}{-7}}$ & $66.9$ & ${\sci{9.76}{-8}}$ \\ \midrule[1pt]
		Risk (MW) & \multicolumn{2}{c}{$\mathbf{\sci{1.76}{-1}}$} & \multicolumn{2}{c}{$\sci{1.60}{-2}$}  \\
		\bottomrule[1.5pt]
	\end{tabularx}
\end{table}

It should be noted from Table. \ref{tab:npcc_risk_summer_path1_rates} that the overheat of transmission line conductors have a significant effect towards increasing the outage rate. Since the weather condition on Aug. 11 and on Aug. 22 are not exactly the same, to achieve fair comparison, we also compare the outage rates of same line, at same time interval but with different temperatures, as shown in Table. \ref{tab:npcc_risk_summer_path1_0811}. It can be seen that under the same environmental conditions, the higher conductor temperature brings a higher rate of line outage. This case verifies that the line overheat caused by the power flow re-distribution is a source of interdependency of cascading outages.

With the approximate analytical solution of TLTE as a function of transmission line current in \cite{yao2017efficient}, the sensitivity of outage probability to system states can also be estimated, which could facilitate the derivation of strategies for reducing cascading outage risk \cite{yao2017reduction}.

\begin{table}[htb]
	\centering
	\caption{Outage rates and temperature on August 11, 2016}
	\label{tab:npcc_risk_summer_path1_0811}
	\begin{tabularx}{\linewidth}{p{1.5cm}@{}*4{>{\centering\arraybackslash}X}@{}}
		\toprule[1.5pt] 
		\centering\multirow{2}*{Outage} & \multicolumn{2}{c}{In path C} &
		\multicolumn{2}{c}{In other paths} \\ \cline{2-5}
		& Max $T_c$ (${}^{\circ}$C) & Outage rate (s${}^{-1}$) & 
		Max $T_c$ (${}^{\circ}$C) & Outage rate (s${}^{-1}$) \\ \midrule[1pt]
		88          & $104.4$ & $\mathbf{\sci{9.86}{-7}}$ & $52.8$ & ${\sci{8.97}{-8}}$ \\
		183         & $102.5$ & $\mathbf{\sci{3.30}{-6}}$ & $47.1$ & ${\sci{3.29}{-7}}$ \\
		190         & $92.3$  & $\mathbf{\sci{5.79}{-7}}$ & $45.0$ & ${\sci{1.29}{-7}}$ \\
		\bottomrule[1.5pt]
	\end{tabularx}
\end{table}

\subsection{Two-stage risk assessment}
We test the effectiveness and performance of the two-stage risk assessment based on MT search. The NWP outreach of HRRR (high-resolution rapid refresh) on Aug. 11 is 15 hours\cite{benjamin2016north}. The forecast for 13:00-16:00 EDT Aug. 11 can be acquired from the data of 01:00 EDT Aug. 11 (available at around 02:00 EDT). We use the NWP data of 01:00 EDT to conduct risk assessment for the time interval 13:00-16:00 EDT first, and then update the result in the second stage with NWP data of 11:00 EDT (available at around 12:00 EDT).


\begin{table}[htb]
	\centering
	\caption{Computational configurations of MT risk assessment}
	\label{tab:npcc_risk_config}
	\begin{tabularx}{\linewidth}{p{1.5cm}@{}*2{>{\centering\arraybackslash}X}@{}}
		\toprule[1.5pt]
		Config.        & Offline   & Online   \\ \midrule[1pt]
		Config. 1      & --        & 12:00 MT search using 11:00 EDT NWP data     \\
		Config. 2      & 1200 MT search using 01:00 EDT NWP data     & Update results with 11:00 EDT NWP data     \\
		Config. 3      & 800 MT search using 01:00 EDT NWP data     & Update offline results \& continued 400 MT search with 11:00 EDT NWP data     \\
		\bottomrule[1.5pt]
	\end{tabularx}
\end{table}

\begin{table}[htb]
	\centering
	\caption{Risk and computation time under different configurations}
	\label{tab:npcc_risk_compare_two_stage}
	\begin{tabularx}{\linewidth}{p{0.1cm}@{}*4{>{\centering\arraybackslash}X}@{}}
		\toprule[1.5pt]
		& \centering\multirow{2}*{Configuration}  & \centering\multirow{2}*{Risk (MW)} & \multicolumn{2}{c}{Computation time (s)}\\ \cline{4-5}
		&      &    & Offline & Online\\ \midrule[1pt]
		& Config. 1      & 18.23     & --      & 2226.04  \\
		& Config. 2      & 17.69     & 2257.94 & 763.94   \\
		& Config. 3      & 18.22     & 1752.49 & 1075.30  \\
		\bottomrule[1.5pt]
	\end{tabularx}
\end{table}

\begin{figure}[htb]
	\centering
	\includegraphics[width=6cm]{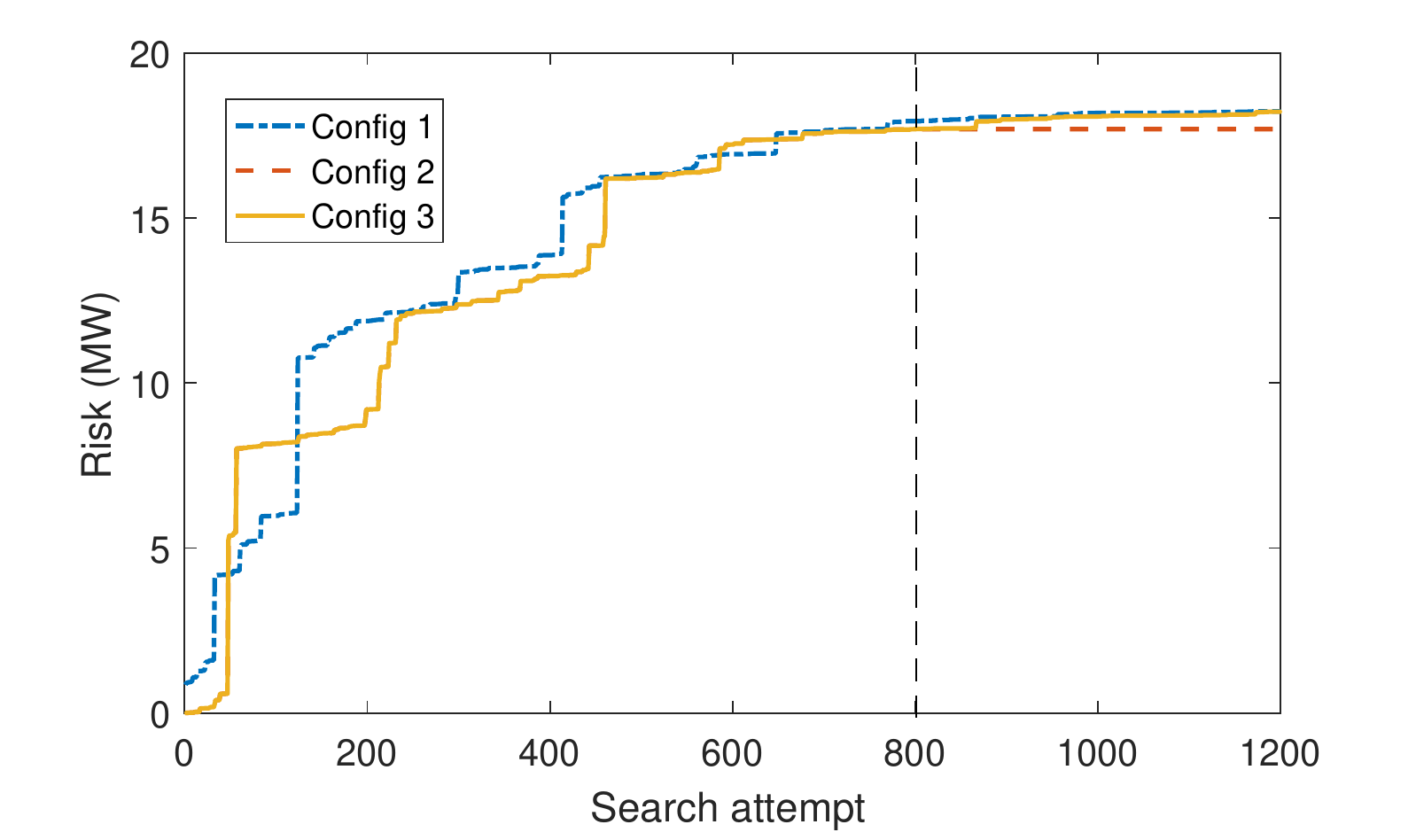}\\
	\caption{MT risk assessment results under different configurations}\label{fig:160811_mt_risk}
\end{figure}

We compare 3 configurations of risk assessment listed in Table. \ref{tab:npcc_risk_config}. The config. 1 completely relies on the online computation. Config. 2 conducts full risk assessment in the offline stage and only updates results using the latest NWP data. Config. 3 is the proposed two-stage method, which assigns most risk assessment offline and update the result online with latest weather data, and then the risk assessment is continued to search for emerging risk brought by the updated weather condition. Table. \ref{tab:npcc_risk_compare_two_stage} shows the results of risk assessment and the computational time under each configuration. Compared with Config. 1, the Config. 2 reduces online computation time by about 70\%, but risk is underestimated by about 3\%. The underestimation of risk is caused by the MT search using old weather data, which does not fully reflect the risk in the actual weather conditions. Such shortcoming is overcome by the two-stage method. In Config. 3, the final risk assessment result is almost the same as Config. 1, and the online computation time is reduced by more than 50\%. From the convergence profiles of the 3 configurations of MT search shown in Fig. \ref{fig:160811_mt_risk}, the Config.2 and Config. 3 are the same in the first 800 MT search attempts, while after 800 MT search attempts, the Config. 2 finds almost no new risk, while Config. 3 manages to search out about 3\% risk in the online stage with updated weather data. This case verifies the advantage of two-stage risk assessment in both accuracy and efficiency for online assessment.

\section{Conclusion}
This paper proposes a method for the risk assessment of weather-related cascading outages in power systems. First according to the outage mechanisms, a generic outage rate model of transmission lines dependent on weather condition and conductor temperature is proposed, and then based on the polynomial approximation of outage rate and the analytical solution of transmission line temperature evolution (TLTE), the analytical expression of outage probability of transmission line is derived, which facilitates efficient risk assessment of weather-related cascading outages. Based on the modeling of physical outage mechanisms, i.e. weather-induced outages, outages caused by sagging into vegetation and by damage of overheat, the outage rates of transmission lines can be estimated with given weather conditions, historical outage records and conductor parameters. With the weather-dependent outage model, a two-stage risk assessment method based on Markovian tree (MT) search is proposed. The method consists of offline full assessment with early weather prediction data and efficient online update of risk assessment and continued supplementary risk assessment with imminent weather prediction data, which has advantage both in accuracy and efficiency.

The method is tested on an NPCC 140-bus test system model with geographical positions. The tests on winter and summer scenarios verify that the proposed method can reflect the threat of adverse weather conditions on the power systems. Also the advantage of the two-stage risk assessment in accuracy and efficiency is demonstrated.


%




\ifCLASSOPTIONcaptionsoff
  \newpage
\fi



%

\bibliographystyle{IEEEtran}
\bibliography{bib/refs}

%








\end{document}